# Angiogenesis regulators as a possible key to accelerated growth of secondary tumors following primary tumor resection

Irina Kareva[1,2*]

[1] Mathematical, Computational and Modeling Sciences Center, Arizona State University, Tempe, AZ, 85287
[2] EMD Serono Research Center, Billerica, MA, 01821
[*]ikareva@asu.edu

**Abstract**

Resection of primary tumors is often followed by accelerated growth of metastases. Here we propose that this effect may be due to the fact that resection of primary tumor results in a decrease in the total systemic amount of angiogenesis stimulators, such as VEGF and bFGF. This in turn causes decrease in the systemic level of angiogenesis inhibitors, such as PF-4 and TSP-1, which at least temporarily relieves inhibition of secondary tumors, allowing them to grow. This construct is predicated on the notion that systemic level of angiogenesis inhibitors is regulated by the systemic level of angiogenesis stimulators, as the host is trying to maintain the homeostatic balance of stimulators to inhibitors in the body. We evaluate this hypothesis using a conceptual mathematical model and show that indeed, this mechanism can explain accelerated growth of secondary tumors following resection of a primary tumor. We also show that there exists a tradeoff between time of surgery and time to onset of metastatic growth. We conclude with a discussion of possible therapeutic approaches that may counteract this effect and reduce metastatic recurrences after surgery.

**Introduction**

Metastases are a primary cause of death of cancer. It is likely that secondary tumors have been seeded early in tumor development (1). Furthermore, there exist numerous observations that surgical removal of primary tumors can lead to increased growth of secondary tumors (2–6), suggesting that primary tumors may have had some inhibitory effect on metastases. This phenomenon has been observed in colorectal cancer (5), non-small-cell lung cancer (7), among other cancer types.

A number of hypotheses have been proposed to explain this observation, including monopolization of resources by the primary tumor, depriving secondary tumors of necessary nutrients; immune stimulation, induced by the primary tumor, which causes cross-immunity and thus suppression of secondary tumors (8–10); intrinsic population heterogeneity which dictates time to escape from dormancy regardless of extrinsic events (11), among others. Here we explore another possibility, which focuses on the systemic response of the host to tumor-induced production of angiogenesis stimulators, and on the long-term effects of the altered balance between angiogenesis stimulators and inhibitors in the presence of more than one tumor in the host's body.

Tumors cannot grow beyond 1-2mm$^3$ without recruiting their own blood supply, a process that has been known as 'angiogenic switch' (12–14). The ability of a tumors to recruit their own blood supply relies in part on a tumor-induced shift in the balance between pro-angiogenic and anti-angiogenic proteins away from the normal physiological 'off' state (15). This is achieved when the pro-angiogenic effects of continuously produced angiogenesis stimulators, such as vascular endothelial growth factor (VEGF) and basic fibroblast growth factor (bFGF), overwhelm the inhibitory effects of angiogenesis inhibitors, such as platelet factor 4 (PF-4), thrombospondin-1 (TSP-1), angiostatin and endostatin (15–18).

Angiogenesis stimulators, such as VEGF, induce formation of early vascular sprouts, or tip cells (19). Sprout growth and lumenization (formation or stalk cells) is facilitated by factors, such as bFGF (20), henceforth for brevity referred to as blood vessel stabilizers. All of these angiogenesis regulators bind to the same binding cite, namely, heparan sulfate (21–24). Angiogenesis inhibitors facilitate termination of neovascularization either by outcompeting angiogenesis stimulators due to high affinity for heparin, as is the case for PF-4 (25) or via repressing anti-apoptosis genes and inhibiting factors involved in cell cycle arrest, causing proliferating cells to become apoptotic, as is the case for endostatin and angiostatin (26,27).

In (28), we have investigated the mechanism of normal and pathological blood vessel formation as mitigated by aforementioned angiogenesis regulators. We used an experimentally validated agent-based model to describe the process of angiogenesis regulator mediated neovascularization that can become unrestrained if angiogenesis inhibitors become outcompeted and thus unable to function. This becomes possible if there exists an excessive inflow of angiogenesis stimulators and blood vessel stabilizers, which can outcompete angiogenesis inhibitors for binding cites (specifically heparan sulfate), thus precluding termination of angiogenesis and facilitating continuous formation of new vessels.

In follow-up work in (18), we built a mathematical model that demonstrated that tumor-induced production of angiogenesis stimulators and blood vessel stabilizers, which outcompete angiogenic inhibitors in the tumor microenvironment, results in accumulation of sufficient neovasculature to permit self-supporting tumor growth via a positive feedback loop. Through varying parameters governing the degree of tumor-induced production of angiogenesis stimulators and blood vessel stabilizers, we were able to qualitatively replicate experimentally observed growth curves for both dormant and actively growing tumors of breast cancer and liposarcoma. In fact, variation of only two parameters was sufficient to replicate any experimentally observed time to angiogenic switch in the available data. The model thus provided a possible mechanistic explanation for escape from dormancy of primary tumors.

Here, we build on this work to investigate a hypothesis about mechanisms that can underlie escape from metastatic tumor dormancy, and specifically, the observation that secondary (metastatic) tumors may experience accelerated growth following resection of a primary tumor.

Many tumors are characterized by increased production of angiogenesis stimulators, such as VEGF (29–31). Furthermore, it has been observed that the level of inhibitors in the tumor host is also elevated, production of which has been attributed to the tumors themselves, suggesting a degree of self-inhibition (32–34). This hypothesis was mathematically investigated by Benzekry et al. (35), where the authors identified ranges of parameter values that allow for an organism-level homeostatic steady state in total tumor burden through mutual angio-inhibitory interactions within a population of tumor lesions that could yield global dormancy.

Here we propose an alternative model to explain a possible mechanism, whereby secondary tumors might undergo accelerated growth following removal of the primary tumor. Specifically, a possible explanation for the presence of high levels of inhibitors is the normal tissues' response to increased presence of angiogenesis stimulators. This increase would occur in order to restore the homeostatic ratio of angiogenesis stimulators to inhibitors within the host, returning it to the physiologically normal 'off' state. Gonzalez et al. have shown that the ratio of inhibitors to stimulators, such as TSP-1 to VEGF, in serum and in serum minus plasma, is twice to three times as high in healthy subjects as it is in cancer patients (36). Furthermore, some preliminary data suggest that injection of VEGF into mice resulted in increased levels of angiogenesis inhibitors (Giannoula Klement, personal communication), suggesting that increase in level of stimulators would be followed by increase in systemic levels of angiogenesis inhibitors, as the body would attempt to suppress unnecessary neovascularization.

If the body indeed responds with increasing production of angiogenesis inhibitors in an attempt to compensate for increased stimulator production to restore the state of homeostasis, one can surmise of the following scenario:

Assume the presence in the host of two tumors, a larger primary tumor and a much smaller secondary tumor. Both of them are promoting to varying degrees the secretion of some angiogenesis stimulators. At some point this may become sufficient to trigger the body to commence production of additional angiogenesis inhibitors to restore homeostasis. It is possible that this increase in the systemic level of angiogenesis inhibitors may be sufficient to suppress the smaller tumor.

Surgical removal of the primary tumor would also remove the source of a large amount of angiogenesis stimulators, effectively reducing or even eliminating the need to further produce additional inhibitors in order to maintain homeostasis. Therefore, the amount of inhibitors would decrease, giving the secondary tumor an opportunity to build up vasculature and grow, perhaps to the point when newly activated systemic production of angiogenesis inhibitors can no longer keep up. At this point, the positive feedback loop of tumor-induced production of self-supporting vasculature would have progressed too far. This mechanism, which is based on normal physiological responses and the body's attempts to restore homeostasis, could provide a possible explanation for why removal of primary tumors could result in accelerated growth of secondary tumors.

This scenario requires the following assumptions to hold:

1) Levels of angiogenesis inhibitors are responsive to the current levels of angiogenesis stimulators, i.e., when the concentration of angiogenesis stimulators increases, the host's body tries to restore homeostatic ratio of angiogenesis regulators by increasing the levels of inhibitors.

2) There exists a threshold of sensitivity of inhibitors to the presence of angiogenesis stimulators in the body of the host, such that when the total amount of stimulators is above this threshold, secretion of inhibitors increases in order to compensate for stimulator increase; below this threshold, the amount of inhibitors returns to homeostatic levels (15).

This scenario is summarized in Figure 1.

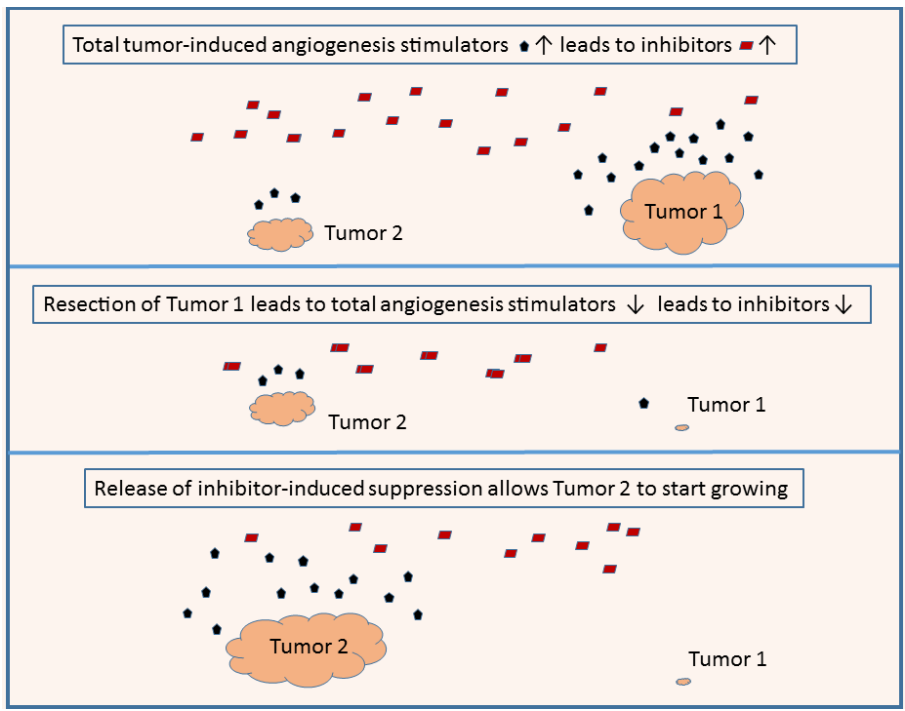

**Figure 1**. Proposed scenario of mechanisms underlying accelerated growth of secondary tumors following resection of a primary tumor. Suppose the host is harboring two tumors. Cumulative amount of angiogenesis regulators could trigger a natural response, where the host would attempt to restore homeostasis of angiogenesis regulators by producing additional inhibitors. Resection of one of the tumors would reduce the total number of angiogenesis stimulators, causing responsive decrease in angiogenesis inhibitors, thus allowing secondary tumor to progress. This Figure is used with permission from (37).

This hypothesis is supported by several studies, where the authors evaluated the levels of angiogenesis stimulators and inhibitors before and after surgery. Dipok et al. (38) reported that significant decreases in serum endostatin and bFGF were observed in postoperative samples of patients with hepatocellular carcinoma, compared to the preoperative values. Similar results were observed by Feldman et al. (39) for patients with colorectal cancer. The authors reported that plasma endostatin levels were significantly higher in the 30 patients compared to controls before surgery. However, none of the patients who remained progression free had elevated endostatin levels at follow up. Several studies also report that elevated post-operative endostatin levels were associated with poor prognosis for patients with advanced stage nasopharygeal carcinoma (40) and for human malignant gliomas (41), potentially suggesting presence of metastatic tumors.

Here, we introduce a proof of concept model, which captures the scenario described above. We identify the necessary assumptions and perform sensitivity analysis in order to identify key parameters that may be driving system dynamics. We demonstrate that indeed, surgical removal of the primary tumor can lead to a cascade of responses that would result in releasing the inhibitory effect on the secondary tumor(s), providing a possible explanation for accelerated growth of metastases following resection of the primary tumor. We conclude with a discussion of therapeutic options.

### Model description

In order to evaluate the hypothesis that the host's response to removal of primary tumor can relieve inhibition of a secondary tumor within this host, we propose the following mathematical model, which takes into account the dynamical activity over time of the following variables: primary tumor $T_1(t)$, its dynamic vasculature $V_1(t)$, tumor-induced angiogenesis stimulators, such as VEGF, described by $S_1(t)$; secondary tumor $T_2(t)$, its dynamic vasculature $V_2(t)$, tumor-induced angiogenesis stimulators $S_2(t)$; and the shared pool of angiogenesis inhibitors, such as PF-4 and TSP-1, described by $I(t)$.

We expect that given our understanding of the system, we will see the dynamics as depicted on Figure 2, where without surgery the overall levels of angiogenesis stimulators will remain high, causing high levels of angiogenesis inhibitors and consequent suppression of the secondary tumor. However, in the event of a surgery, amount of inhibitors will decrease, causing increase in the growth of secondary tumors.

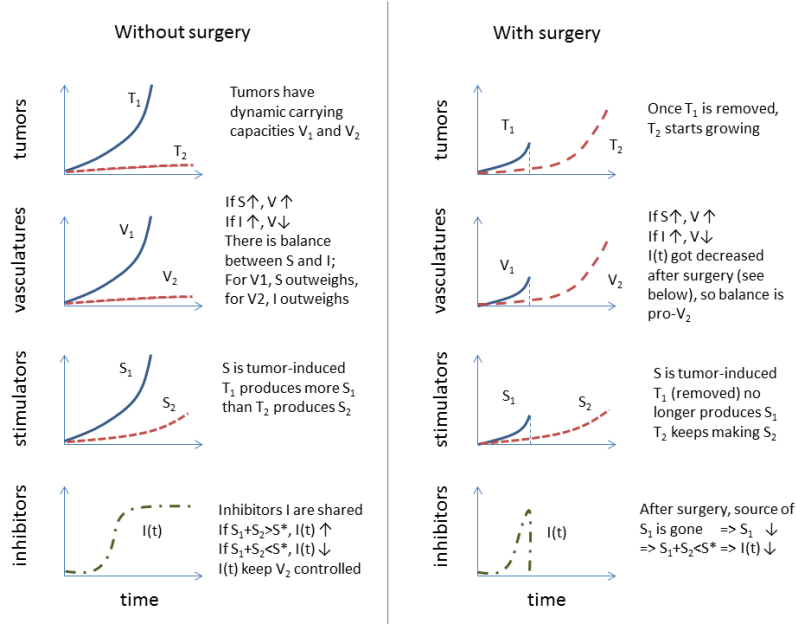

**Figure 2.** Schematic representation of the anticipated dynamics of tumor, neovasculature, stimulators and inhibitors with and without surgery. We expect that without surgery, the level of angiogenesis inhibitors will remain high, inhibiting secondary tumor growth. However, we predict that after resecting the primary tumor, the level of inhibitors will decrease due to decrease of overall stimulator levels, causing accelerated growth of the secondary tumor.

We propose that the change in volume of the primary tumor $T_1(t)$ is described by an Allee-type growth function with a dynamic carrying capacity

$$\frac{dT_1}{dt} = \underbrace{\lambda_1 T_1(t)(\frac{T_1(t)}{m_1}-1)(1-\frac{T_1(t)}{k+V_1(t)})}_{\text{Allee growth with vasculature dependent dynamic carrying capacity}}$$

, where $\lambda_1$ is the growth rate of the tumor cells, $k$ is tumor carrying capacity in the absence of neovascularization (the maximum size that a dormant non-vascularized tumor can reach), and $V_1(t)$ is the volume of new vasculature, formed as a result of the activity of angiogenesis stimulators and inhibitors. Parameter $m_1 \neq 0$ describes the threshold of viability of a resected tumor; when $T_1(t) > m_1$, the tumor grows up to the dynamic carrying capacity $V_1(t)$. The threshold $m_1$ reflects an observation that in a 3-dimensional culture, depending on the environment and the surrounding signaling, at low concentrations malignant cells can sometimes revert to non-malignant phenotype (42–44). This suggests existence of some threshold of viability, which is accounted for by parameter $m_1$.

In our previous work (18) we looked at three classes of angiogenesis regulators: stimulators, such as VEGF, blood vessel stabilizers, such as bFGF and PDGF, and angiogenesis inhibitors, such as PF-4 and TSP-1. For the purposes of the question considered here, we can reduce the number of variables without loss of qualitative behavior, and combine stimulators and stabilizers into a single class of "angiogenesis stimulators" $S_1(t)$, which act to promote formation of new blood vessels.

Here, we will describe the dynamics of tumor-induced vasculature, as well as tumor-induced production of angiogenesis stimulators, and the systemic response to them of angiogenesis inhibitors. That is, we are looking at levels of angiogenesis regulators, as well as neovasculature created above the body's normal baseline. For this reason, in the absence of a tumor, the level of tumor-induced stimulators will tend to zero. In the presence of a tumor, there exists a tumor-induced inflow of angiogenesis stimulators, described by the ratio dependent functional form $r_1 S_1(t) \frac{T_1(t)/S_1(t)}{1+T_1(t)/S_1(t)} = r_1 \frac{S_1(t)T_1(t)}{S_1(t)+T_1(t)}$, where production of additional stimulators is proportional to the ratio of $T_1(t)/S_1(t)$. This functional form allows additional tumor-induced production of stimulators to be dependent both on the current amount of stimulators, and on the volume of tumor present. We also assume there to be some natural clearance rate of angiogenesis stimulators, described by $\delta S_1(t)$.

The equation for angiogenesis inhibitors $I(t)$ is derived in the following way. Angiogenesis stimulators $I(t)$ will increase if the total sum of angiogenesis stimulators $S_1(t) + S_2(t)$ is greater than some threshold value $S*$ up to some carrying capacity $I_{max}$, and decrease to zero if the sum of total stimulators $S_1 + S_2$ is less than $S*$. As with stimulators, we track only the increases in inhibitors that result from the presence of the tumor(s). This effect is captured by an Allee-effect type functional form $r_I I(t)(\frac{S_1(t)+S_2(t)}{S*}-1)(1-\frac{I(t)}{I_{max}})$. Angiogenesis inhibitors are also cleared at some constant rate, described by the term $\delta_3 I(t)$. Notably, as was mentioned above, there exist two classes of mechanisms whereby angiogenesis inhibitors contribute to angiogenesis inhibition, namely, by directly outcompeting angiogenesis stimulators, or by interfering with endothelial cell proliferation and inducing apoptosis. Here, we focus on the second mechanism. A detailed investigation of the effects of competition for binding cites between angiogenesis stimulators and inhibitors can be found in (28,45).

Finally, the equations for the dynamics of tumor-induced vasculature are derived as follows. Increase in vasculature is proportional to the presence of angiogenesis stimulators $\gamma_1 S_1(t)$, and decrease can occur either naturally at some low rate $\xi_1 V_1(t)$, or due to angiogenesis inhibitors, which we propose occurs at a rate $\frac{v_1 I(t)}{1+I(t)}$. The balance between stimulators and inhibitors will determine, whether tumor-induced vasculature will increase or not. Equations describing the dynamics of the secondary tumor and the corresponding vasculature volume and amount of angiogenesis stimulator, are derived in the same way, varying only in parameter values.

Noticeably, as was mentioned above, mechanism of action of angiogenesis inhibitors like angiostatin and endostatin involve increased apoptosis, which may include some part of the existing neovacsulature. This is observed during normal blood vessel formation, when the vasculature is initially overbuilt and then pruned, decreasing in size to a stable level. As one can see from the equations, in the absence of a tumor, $S(t) \to 0$, and therefore $I(t) \to 0$, and consequently $V(t) \to 0$. However, there may be a short transient period, when the value of $V(t)$ briefly becomes negative, indicating excessive inhibitor-induced apoptosis that affects the state of the baseline vasculature. This is consistent with current understanding of the biology of angiogenesis regulators. (This also suggests that a secondary tumor could at least transiently suppress the growth of the primary tumor not only through stimulating system-wide production of angiogenesis inhibitors that would outcompete the stimulators, but also through at least temporary degradation of the vasculature of the secondary tumor, whose stimulator activity does not compensate quickly enough for the systemic increase in the level of angiogenesis inhibitors.)

Taking into account all of these considerations, we obtain the following system of equations:

$$\underbrace{\frac{dT_1(t)}{dt}}_{\text{tumor growth}} = \underbrace{\lambda_1 T_1(t)(\frac{T_1(t)}{m_1}-1)(1-\frac{T_1(t)}{k+V_1(t)})}_{\text{Allee growth with vasculature dependent dynamic carrying capacity}} \qquad \frac{dT_2(t)}{dt} = \lambda_2 T_2(t)(\frac{T_2(t)}{m_2}-1)(1-\frac{T_2(t)}{k+V_2(t)})$$

$$\underbrace{\frac{dV_1(t)}{dt}}_{\text{neovasculature}} = \underbrace{\gamma_1 S_1(t)}_{\substack{\text{increase due to}\\\text{angiogenesis}\\\text{stimulators}}} - \underbrace{\xi_1 V_1(t)}_{\substack{\text{natural}\\\text{decay}}} - \underbrace{\frac{\nu_1 I(t)}{1+I(t)}}_{\substack{\text{inhibitor-induced}\\\text{vasculature decay}}} \qquad \frac{dV_2(t)}{dt} = \gamma_2 S_2(t) - \xi_2 V_2(t) - \frac{\nu_2 I(t)}{1+I(t)} \qquad (1)$$

$$\underbrace{\frac{dS_1(t)}{dt}}_{\substack{\text{angiogenesis}\\\text{stimulators}}} = \underbrace{\frac{r_1 S_1(t) T_1(t)}{S_1(t)+T_1(t)}}_{\substack{\text{tumor-induced production}\\\text{of angiogenesis stimulators}}} - \underbrace{\delta_1 S_1(t)}_{\text{natural decay}} \qquad \frac{dS_2(t)}{dt} = \frac{r_2 S_2(t) T_2(t)}{S_2(t)+T_2(t)} - \delta_2 S_2(t)$$

$$\underbrace{\frac{dI}{dt}}_{\substack{\text{angiogenesis}\\\text{inhibitors}}} = \underbrace{r_I I(t)(\frac{S_1(t)+S_2(t)}{S^*}-1)(1-\frac{I(t)}{I_{max}})}_{\substack{\text{Allee affect type growth: } I(t) \text{ increases if } S_1(t)+S_2(t)>S^*,\\\text{up to carrying capacity } I_{max}, \text{ and decreases if } S_1(t)+S_2(t)<S^*.}} - \delta_3 I(t).$$

All of the variables and parameters for System (1), as well as sample parameter values and initial conditions, are summarized in Table 1.

**Table 1. Variables and parameters for System (1).**

| Parameter/ Variable | Description | Sample value | Units |
|---|---|---|---|
| $T_1(t)$, $T_2(t)$ | Primary and secondary tumor volumes | $T_1(0)=1$ $T_2(0)=1$ | Volume (i.e. mm$^3$) |
| $S_1(t), S_2(t)$ | Amount of angiogenesis stimulators produced by respective tumors | $S_1(0)=0.1$ $S_2(0)=0.1$ | Volume |
| $V_1(t), V_2(t)$ | Tumor(s)-induced vasculature | $V_1(0)=0.01$ $V_2(0)=0.01$ | Volume |
| $I(t)$ | Angiogenesis inhibitors (TSP-1) | $I(0)=3$ | Volume |
| $\lambda_1, \lambda_2$ | Tumor(s) growth rates | 0.04 | 1/time (i.e. day) |
| $m_1, m_2$ | Threshold volume of tumor survival | 0.7 | Volume |
| $k$ | Maximum volume of avascular tumor growth | 1 | Volume |
| $\gamma_1$ | Rate of vasculature growth for $T_1$ | 2.7 | 1/time |
| $\gamma_2$ | Rate of vasculature growth for $T_2$ | 0.5 | 1/time |
| $\xi_1, \xi_2$ | Rate of natural vasculature decay | 0.09 | 1/time |
| $\nu_1, \nu_2$ | Rate of $I(t)$-induced vasculature inhibition | 0.3 | 1/time |
| $r_1$ | Rate of tumor-induced production of $S_1$ | 0.17 | 1/time |
| $r_2$ | Rate of tumor-induced production of $S_2$ | 0.11 | 1/time |
| $\delta_1, \delta_2$ | Stimulator (VEGF) clearance rate | 0.07 | 1/time |
| $\delta_3$ | Inhibitor (i.e., PF-4) clearance rate | 0.01 | 1/time |
| $r_I$ | Rate of increased inhibitor production in response to cumulative $S_1+S_2$ | 0.1 | 1/time |
| $S^*$ | Inhibitor sensitivity threshold to cumulative $S_1+S_2$ | 3 | Volume |
| $I_{max}$ | Carrying capacity for angiogenesis inhibitors | 100 | Volume |

**Sensitivity Analysis**

Due to lack of consistent experimental data for the proposed proof-of-concept model, most parameter values were chosen in such a way as to capture the theoretically predicted dynamics. That is, our goal is to evaluate whether the proposed set of assumptions would be sufficient to reproduce the qualitative pattern of expected behavior. Despite this limitation, we can use sensitivity analysis to discern relative importance of various parameters in the system over time.

Within each parameter there can be significant variability. Moreover, low dimensionality of the proposed model may result in compression of several parameter values into the same parameter, introducing further variability. In order to evaluate, which parameters in System (1) would have the largest impact on the overall system dynamics at various time points, we perturbed each parameter by a uniformly distributed random variable within the range of 15% of the initial parameter value. The sensitivity indices, which are defined as fractions of total output variance generated by the uncertainty in the respective parameter value, were calculated using the Fourier Amplitude Sensitivity Test (FAST) method (46), an approach that allows investigating the effect of large, concurrent perturbations in model parameters. Larger values of sensitivity indices indicate larger sensitivity to parameter perturbation at each time point evaluated.

The results of the sensitivity analysis performed on System (1) are summarized in Figure 3. Changes over time in the values of the first order sensitivity indices, which reflect sensitivity of each parameter to perturbation, are plotted in Figure 3a. Due to large variations in relative sensitivities of different parameters, the results are additionally presented in Figure 3b-d, where sensitivity indices of parameters pertaining to tumor growth are presented in Figure 3b, sensitivity indices of parameters pertaining to vasculature growth are reported in Figure 3c, and sensitivity indices of parameters pertaining to growth of stimulators are presented in Figure 3d. Only parameters with relative sensitivity indices over 0.01% at any time point are reported.

Seven parameters were identified as sensitive to perturbation (Figure 3). Parameters pertaining to tumor growth (Figure 3b) and vasculature growth (Figure 3c) have the largest impact in the beginning of system development, in the time up until $t \approx 150$. However, as the tumors grow, the parameters that influence the dynamics the most are parameters pertaining to the dynamics of angiogenesis stimulators (Figure 3d), emphasizing the importance of angiogenesis regulators with regards to tumor dynamics.

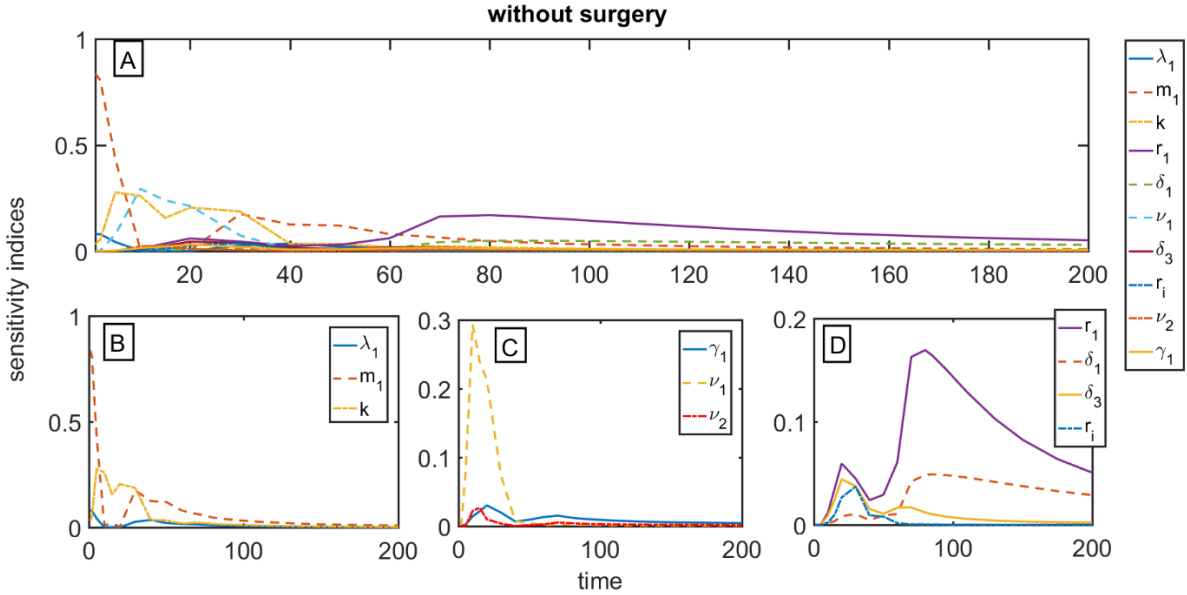

**Figure 3.** First order sensitivity indices for parameters of System (1), obtained using the FAST method. Only indices that are over 0.01% are reported. (A) All of the identified parameters that are sensitive to perturbation; (B) sensitive parameters that affect tumor growth; (C) sensitive parameters that affect the growth of neovasculature; (D) sensitive parameters that affect the dynamics of tumor-induced angiogenesis stimulators. All parameters are taken from Table 1.

Out of the seven parameters that were identified using the FAST method (Figure 3a), we will vary only $\gamma_1$ and $r_1$, which describe rate of stimulator-dependent vasculature formation, and rate of tumor-induced production of stimulators, respectively. In the next section, these two parameters will be the only ones that differ between $T_1$ and $T_2$. This will allow us to narrow our investigation down to effects of tumor-induced vasculature formation and not on individual intrinsic tumor properties.

**Results**

In order to evaluate our hypothesis that resection of a primary tumor may lead to decrease in systemic level of inhibitors, facilitating growth of secondary tumors, we performed a series of numerical experiments, where we simulated resection of the primary tumor, and evaluated how quickly secondary tumors start growing compared to scenario with no tumor resection. All the parameter values were taken from Table 1.

As one can see in Figure 4, without resection, the size of the primary tumor $T_1$ predictably keeps increasing. However, with surgery performed at t=100, removing 99.99% of the primary tumor, causing $T_1$ size to decrease below threshold $m_1$, secondary tumor $T_2$ starts growing sooner than it would have without surgery.

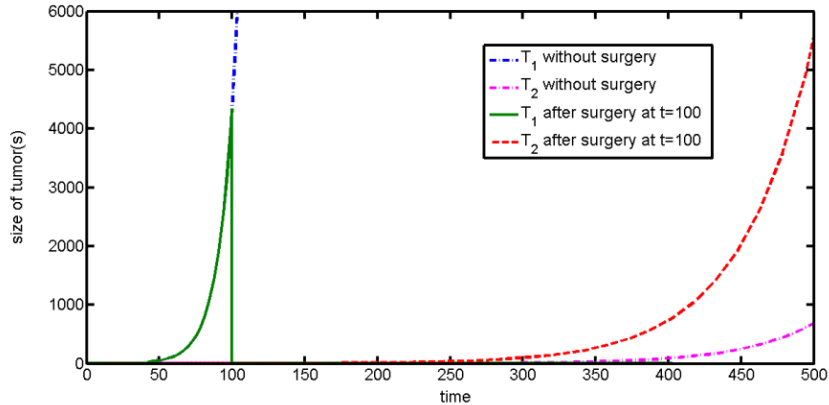

**Figure 4.** Comparison of primary and secondary tumor dynamics without surgery and with surgery at t=100. As one can see, without surgery, the primary tumor keeps growing, but with surgery, secondary tumors start growing sooner.

To better understand this effect, let us evaluate the dynamics of other components of the system, namely, the dynamics of vasculature, angiogenesis stimulators and angiogenesis inhibitors, with and without surgery.

In Figure 5, we can see that predictably, the tumor growth dynamics (Figure 5a) corresponds to the tumor's ability to generate neovasculature (Figure 5b). When the amount of stimulators (Figure 5c) increases beyond threshold $S_1+S_2>S^*$, amount of angiogenesis inhibitors starts rising in response, until they reach their carrying capacity $I_{max}$. This is not sufficient to overcome the driving force of angiogenesis stimulators for the primary tumor $T_1$, but may temporarily be sufficient to halt the growth of the secondary tumor $T_2$.

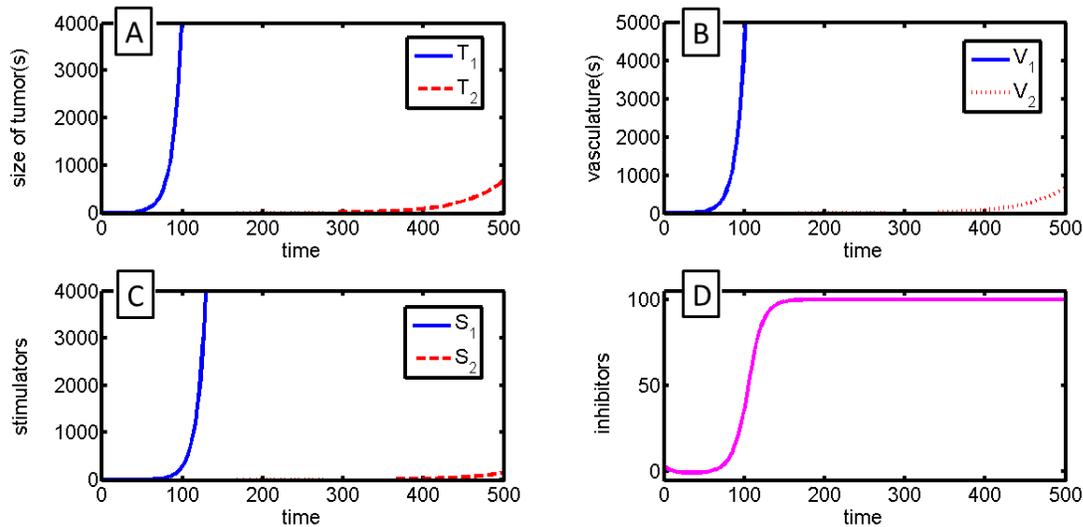

**Figure 5.** Changes over time in tumor size, and corresponding changes in vasculature, angiogenesis stimulators and inhibitors, in a scenario without surgery. (A) Size of primary tumor $T_1$ and (B) corresponding vasculature continue to increase, since (C) the total amount of stimulators remains high, causing (D) total level of inhibitors to remain elevated, delaying growth of the secondary tumor $T_2$.

Now let us compare these dynamics to what happens to system components after surgery (Figure 6). Tumor resection is simulated by reducing the size of the tumor, its vasculature, and amount of stimulators, by the same percentage at a chosen resection time. These new values are then used as initial conditions to continue with the simulation. As one can see, resection of the primary tumor $T_1$ at time *t=100* caused a drop in total number of angiogenesis stimulators (Figure 6c), causing a consequent drop in angiogenesis inhibitors (Figure 6d). This break allows the secondary tumor $T_2$ to produce a sufficient number of angiogenesis stimulators (Figure 6c) to construct neovasculature (Figure 6b) before the total number of stimulators becomes large enough to once again trigger increase in angiogenesis inhibitors (Figure 6d).

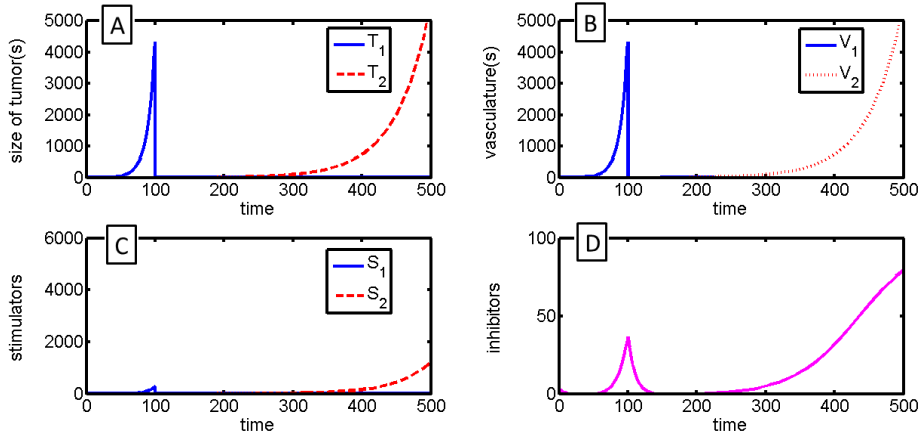

**Figure 6**. Changes over time in tumor sizes, and corresponding changes in vasculature, angiogenesis stimulators and inhibitors, in a scenario with surgery performed to resect the primary tumor at time *t=100*. (A) Resection of primary tumor $T_1$ causes reduction in (B) vasculature and (C) amount of stimulators, in turn causing (D) temporary decrease in amount of inhibitors, facilitating growth of the secondary tumor $T_2$.

It is easy to see from the equation for vasculature dynamics $V'(t)$ why this delay in activation of angiogenesis inhibitors allows building up of sufficient vasculature to overcome the inhibitory effects of *I(t)*. $V'(t) = 0$ when $V^* = \frac{\gamma}{\xi} S(t) - \frac{\nu}{\xi} \frac{I(t)}{1+I(t)}$. As one can see, the dynamics is determined by the balance between $\frac{\gamma}{\xi} S(t)$ and $\frac{\nu}{\xi} \frac{I(t)}{1+I(t)}$, and if the total number of stimulators *S(t)* becomes very large, it will outweigh the inhibitory effects of *I(t)*, pending values of parameters $\gamma$ and $\nu$. The explains why even after the number of inhibitors starts increasing, if the tumor has been able to produce a sufficient number of stimulators, the inhibitors will have little effect.

*Time to surgery and time to metastases*

Finally, we investigated the effects of timing of the surgery on time to appearance of secondary tumors. We simulated surgery at *t=90*, *t=100* and *t=110*. As one can see in Figure 7, secondary tumor volume is offset by the delay in time of surgery. Therefore, there exists a tradeoff: earlier surgery will reduce the size of the primary tumor $T_1$ (Figure 7a), but it will also reduce the delay in time to onset of secondary tumor growth (Figure 7b). Conversely, delaying surgery would allow the primary tumor to reach a larger size, potentially making it unresectable, causing more immediate harm to the patient. Noticeably, within this construct, an important distinction needs to be made: resection of primary tumors does not induce growth of secondary tumors but simply changes time to onset of metastatic growth.

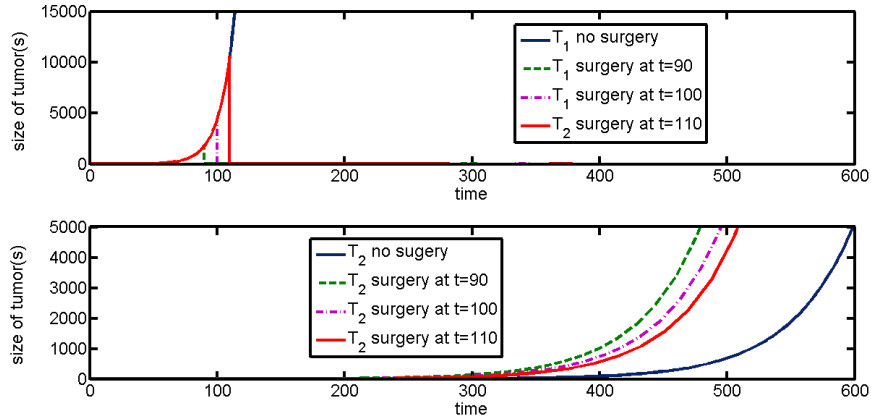

**Figure 7**. Variations in time to appearance of $T_2$ depending on timing of the surgery. Earlier surgery (top panel) results in faster growing metastases (lower panel). However, later resection of the primary tumor allows it to reach larger sizes, potentially making it unresectable and causing more immediate harm to the patient.

In order to further understand how the system changes after surgery compared to before surgery, we performed additional sensitivity analysis to identify, whether and which parameters become more or less important. Results are reported in Figure 8.

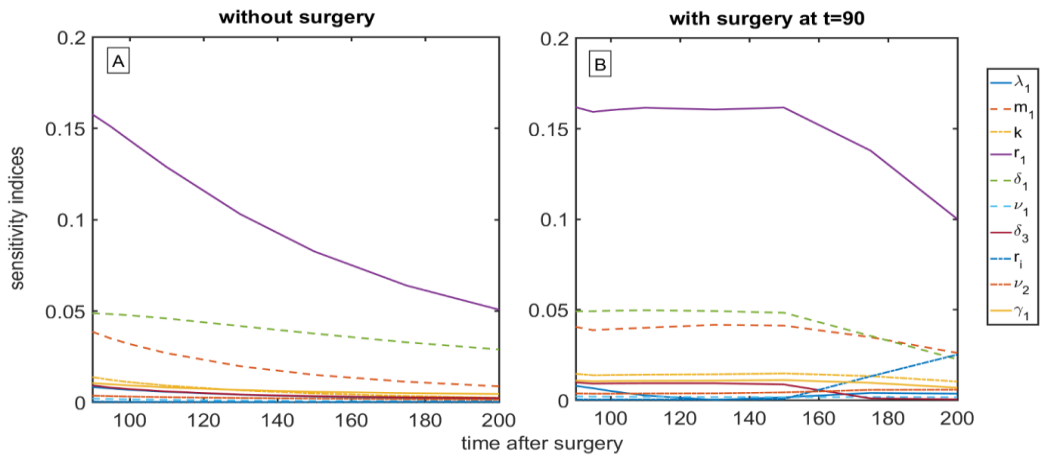

**Figure 8.** Comparison of first order sensitivity indices for parameters of System (1) obtained using the FAST method for (A) simulation of system dynamics without intervention and (B) simulation of system dynamics with surgery at t=90.

Sensitivity indices are predictably identical in the time before surgery, so we report only the changes that occur after surgery. As one can see in Figure 8, parameters $r_1$ and $r_i$ have greater impact on system dynamics after the surgery than in the control case, indicating the high importance of both tumor-induced production of angiogenesis stimulators, and especially whether inhibitors can respond in time to changes in systemic levels of stimulators. The rate at which the host can adapt to the change in angiogenesis stimulators becomes a crucial factor in determining, whether secondary tumors will experience accelerated growth in the time after surgery.

**Discussion**

In this work we propose a theoretical construct that provides a possible explanation for accelerated growth of secondary tumors following resection of a primary tumor. We build on our previous work (18), where we looked at effects of tumor-induced stromal stimulation on time to escape from dormancy. We showed that increased production of tumor-induced stimulators, such as VEGF, bFDF and PDGF, allowed for buildup of vasculature, which eventually initiated a rapid positive-feedback loop of self-supported tumor growth.

Here we expand on that construct to investigate the systemic effects of angiogenesis regulators on growth dynamics of both primary and secondary tumors. We propose that if there exists a systemic response, whereby amount of angiogenesis inhibitors increases in response to elevated levels of angiogenesis stimulators, then resection of a primary tumor will cause an overall drop in total amount of stimulators. This would cause decrease in inhibitors, which may consequently facilitate growth of secondary tumors (Figure 1).

In order to evaluate this hypothesis, we created a proof-of-concept mathematical model, which tracks the dynamics of a primary tumor, a secondary tumor, their respective vasculatures, tumor-induced stimulators, and angiogenesis inhibitors. The dynamics of angiogenesis inhibitors depends on the sum of stimulators produced by both tumors in such a way that if this sum drops below a certain threshold, the amount of inhibitors decreases. We then simulated surgery by reducing the size of the primary tumor by 99.99% and observed the dynamics of secondary tumors. The theoretically predicted growth curves are shown in Figure 2, and the results of the sensitivity analysis, which allowed identifying parameters most sensitive to perturbation, are summarized in Figure 3.

In our numerical simulations, we showed that indeed, this mechanism, if true, can explain accelerated growth of secondary tumors following resection of primary tumor. As one can see in Figure 4, resection of the primary tumor $T_1$ at time $t=100$ reduces time to onset of secondary tumor growth compared to no surgery. In Figures 5 and 6 one can see that this effect indeed comes from temporary reduction in the amount of angiogenesis inhibitors $I(t)$, which allows for secondary tumor to build up sufficient vasculature to overcome the inhibitory effects of $I(t)$.

Next, we investigated the effects of different timing of the surgery on the dynamics of the two tumors. We showed that there exists a tradeoff between surgery time and time to accelerated growth of secondary tumors: earlier surgery results in faster growth of secondary tumors, while later surgery allows the primary tumor to grow to an unresectable size, potentially causing more problems in the short term than secondary tumors would cause in the long term. The importance of systemic response of the host to changes in angiogenesis inhibitors is highlighted by sensitivity analysis performed on System (1) with and without surgery. Results reported in Figure 8 indicate that after surgery, rates of tumor-induced production of angiogenesis stimulators and rate at which the host can respond with production of angiogenesis inhibitors are most important factors in driving system dynamics after surgery.

*Therapeutic implications*

Even though there exists a trade-off between surgery timing and accelerated growth of metastatic tumors, in the presence of a clinically apparent tumor, the question is not whether to resect it or wait, but how to address the danger of potential metastases. In this system, accelerated growth of secondary tumors following the resection of a primary tumor occurs when the total number of angiogenesis stimulators decreases so much as to cause decrease in the number of angiogenesis inhibitors. One possible way to address this issue would be to attempt to maintain the stimulator-inhibitor balance following surgery, either through mimicking the presence of stimulators to maintain elevated levels of inhibitors, or through extrinsically increasing number of inhibitors.

Another, perhaps complementary, approach would be to directly target the source of stimulators after surgery. Targeting the stroma with maintenance therapy, such as for ALL, where a period of high-intensity induction is followed by 2-3 years of lower-dose, higher-frequency therapy (47), or metronomic therapy, which similarly involves administering lower doses of chemotherapeutic agents at more frequent time intervals (48–50), after surgical resection might halt the growth of secondary tumors and reduce the risk of metastatic recurrences.

**Acknowledgements**

The author would like to thank Ben Morin for his invaluable help with coding. The author would also like to thank Georgy Karev for very helpful comments and discussions in model formulation, and Giannoula Klement and Abdo Abou-Slaydi for discussions on tumor biology and angiogenesis. This work was partially supported by NIH/NIGMS R01 GM093050-01A1 (to Giannoula Klement).